\documentclass[12pt]{article}
\usepackage{epsfig}

\newcommand{\beq}{\begin{equation}}
\newcommand{\eeq}{\end{equation}}
\newcommand{\beqa}{\begin{eqnarray}}
\newcommand{\eeqa}{\end{eqnarray}}
\newcommand{\beqar}{\begin{eqnarray*}}
\newcommand{\eeqar}{\end{eqnarray*}}

\newcommand{\eg}{{\it e.g.,}\ }
\newcommand{\ie}{{\it i.e.,}\ }
\newcommand{\labell}[1]{\label{#1}} 
\newcommand{\reef}[1]{(\ref{#1})}
\newcommand\prt{\partial}

\newcommand\cV{{\cal V}}

\newcommand\Tr{{\rm Tr}}

\begin{document}

 \vspace*{1cm}

\begin{center}
{\bf \Large
D-brane-anti-D-brane effective action and \\
brane interaction in open string channel

 }
\vspace*{1cm}

{Mohammad R. Garousi}\\
\vspace*{0.2cm}
{ Department of Physics, Ferdowsi university, P.O. Box 1436, Mashhad, Iran}\\
\vspace*{0.1cm}
{ Institute for Studies in Theoretical Physics and Mathematics
IPM} \\
{P.O. Box 19395-5531, Tehran, Iran}\\
\vspace*{0.4cm}

\vspace{2cm}
ABSTRACT
\end{center}
We construct the effective action of a
$D_p$-brane-anti-$D_p$-brane system by making use of the
non-abelian extension of tachyonic DBI action. We succeed  the
construction by  restricting  the Chan-Paton factors of two
non-BPS $D_p$-branes in the action to the Chan-Paton factors of a
$D_p\bar{D}_p$ system. For the special case that both branes are
coincident, the action reduces to the one proposed by A. Sen.
\\The effective $D_p\bar{D}_p$ potential indicates that when branes separation
is larger than the string length scale, there are two minima in
the tachyon direction. As branes move toward each other under the
gravitational force, the tachyon tunneling from false to true
vacuum may make a bubble formation followed by a classical
evolution of the bubble. On the other hand, when branes separation
is smaller than the string length scale, the potential shows one
maximum and one minimum.  In this case, a homogeneous tachyon
rolling in real time makes an attractive potential for the branes
distance. This classical force is speculated  to  be the
effective force between the two branes.

\vfill \setcounter{page}{0} \setcounter{footnote}{0}
\newpage

\section{The idea} \label{intro}
Study of unstable objects in string theory  might shed new light
in understanding properties of string theory in time-dependent
backgrounds \cite{mgas,asen2,asen3,asen4,nlhl,sen7}. Generally
speaking,  source of instability in these processes  is appearance
of some tachyonic  modes  in the spectrum of these unstable
objects. It  then makes sense to study these objects in a field
theory which includes those modes. In this regard, it has been
shown by A. Sen that an effective action of  Born-Infeld type
proposed in \cite{asen6,mg1,ebmr,jk} can capture many properties
of the decay of non-BPS D$_p$-branes in string theory
\cite{asen2,asen3}. Having an effective action, one may then
study   evolution of these unstable objects in time-dependent
backgrounds. See \cite{Tachyonindustry} for possible cosmological
application of the tachyonic DBI action.

Another unstable object in string theory is parallel D$_p$-brane
anti-D$_p$-brane system (see \eg, \cite{sen2}). Detailed study of
this object reveals   when brane separation is smaller than the
string length scale,  spectrum of this system has two tachyonic
modes \cite{alwis}. These modes however become massive  when
brane separation is larger than the string length scale. The
effective action should then depend on brane separation.  When it
is smaller than the string length scale, the effective action for
${\it dynamics}$ and for ${\it decay}$ of branes should include
the tachyonic modes because these are the most important modes
which rule  the evolution of   the system. On the other hand,
when brane separation is larger than the string length scale, the
effective action which is a low energy effective action,  should
be in terms of massless closed string fields propagating between
the two branes. This action describes properly the ${\it
dynamics}$ of the branes, however, it is  not an appropriate
action for studying ${\it decay}$ of the whole system, \ie bubble
formation followed by a classical evolution of the bubble. This
bubble formation should appear in an effective action which
includes false and true vacuums.

In the literature, there are some proposal for $D_p\bar{D}_p$
effective action when branes are coincident \cite{sen3, kraus}. On
the other hand, when brane separation is larger than the string
length scale,  the low energy gravity effective action is the
Coulomb attraction force due to gravity and RR fields. Making use
of this latter effective action, the cosmology of $D_3\bar{D}_3$
system has been discussed in \cite{dvali}.

In this paper, for ${\it arbitrary}$  branes separation,  we would
like to present an effective action for  parallel $D_p\bar{D}_p$
system in terms of transverse scalars and tachyonic  fields. We
speculate  this effective action should describe the ${\it
dynamics}$ and the ${\it decay}$ of system when branes separation
is smaller than the string length scale, and should describe the
${\it decay}$ of the system when branes separation is larger than
the string length scale.

Our method  for finding this effective action is as follows. One
may try to extract the effective action from the string theory
S-matrix elements, like what has been done for non-BPS $D_p$-brane
case \cite{mg1,mg2}. The tree level world sheet is disk and
various vertex operators are the same as in the non-BPS
$D_p$-brane case. The only difference is that now one has to take
appropriate Chan-Paton factors into account. On the other hand,
the Chan-Paton factors of $D_p\bar{D}_p$ system  are some
subgroup of the Chan-Paton factors of two non-BPS $D_p$-branes,
and the effective action of two non-BPS $D_p$-branes consistent
with string theory S-matrix elements is given by the nonabelian
extension of tachyonic DBI action \cite{mg1}. Restricting the
Chan-Paton factors of  two non-BPS $D_p$-branes in the nonabelian
tachyonic DBI action  to the Chan-Paton factors of a
$D_p\bar{D}_p$ system, we shall find
 effective action of the parallel $D_p\bar{D}_p$
system\footnote{The restriction of Chan-Paton factors may result
from orbifolding type II string theory\cite{sen2}. As pointed out
in \cite{sen2}, the moding out one non-BPS $D_p$-brane in type
IIA/IIB by $(-1)^{F_L^S}$ has a two-fold ambiguity. It may end up
either with a $D_p$-brane or a $\bar{D}_p$-brane in type IIB/IIA.
This gives a four-fold ambiguity for moding out two non-BPS
$D_p$-branes by $(-1)^{F_L^S}$. It may end up with a $D_pD_p$, a
$\bar{D}_pD_p$, a $D_p\bar{D}_p$, or a $\bar{D}_p\bar{D}_p$
system. In general, there should be a $2^N$-fold ambiguity for
moding out $N$ non-BPS $D_p$-branes by $(-1)^{F_L^S}$.}

In the next section we will find the effective action of the
$D_p\bar{D}_p$ system by constraining   the Chan-Paton factors of
different open string fields in the non-abelian tachyonic DBI
action \cite{mg1}. The action includes coupling to gravity
background and world-volume gauge fields.  In section 3, we
simplify the action to include only the real part of tachyon and
the relative distance of the branes. Using  this action,  we then
discuss the ${\it dynamics}$ and the ${\it decay}$ of the branes.

\section{$D_p\bar{D}_p$ effective action}
The proposed  effective action for describing the decay and the
dynamics of a non-BPS D$_p$-brane, and its coupling to gravity
and  world-volume gauge field is given by \cite{mg1,ebmr,jk}:
 \beqa
S&=&-\int d^{p+1}\sigma V(T)
e^{-\Phi}\sqrt{-\det(P[g_{ab}+B_{ab}]+2\pi\alpha'F_{ab}+2
\pi\alpha'\prt_a T\prt_b T)} \,\,,\labell{dbiac2}\eeqa where
$V(T)=T_p(1-\frac{\pi}{2}T^2+O(T^4))$ is the tachyon potential.
Here $g_{ab} , B_{ab},\Phi$ and $A_a$  are the spacetime metric,
antisymmetric Kalb-Ramond tensor, dilaton and the gauge field,
respectively. In above action $P[\cdots]$ is also the pull-back
of the closed string fields. For
 example,  $P[\eta_{ab}]=\eta_{\mu\nu}\prt_a X^{\mu}\prt_b X^{\nu}=
 \eta_{ab}+\prt_a X^i\prt_b X_i$ in the
static gauge\footnote{Our index convention is that
$\mu,\nu,...=0,1,...,9$; $a,b,...=0,1,...,p$  and
$i,j,...=p+1,...,9$.}.

The  kink solution of the equation of motion of tachyon should be
the BPS D$_{p-1}$-brane \cite{sen5}. The tension of the kink is
given by $T_{p-1}= \sqrt{2\pi\alpha'}\int_{-T_0}^{T_0}V(T)dT$
where $T_0$ is the value of the tachyon potential at its minimum.
There are many different tachyon potentials which correctly
reproduce the tension of the BPS brane \cite{jk,sen3,ali}, \ie
$T_{p-1}=\pi\sqrt{2\alpha'}T_p$. One example is the following
potential \cite{kim,nlhl}: \beqa V(T)=\frac{T_p}{\cosh(\sqrt{\pi}
T)}\,\,.\labell{tacpo}\eeqa This  has minimum at $T\rightarrow
\pm\infty$ and behaves as $V(T)\sim e^{-\sqrt{\pi}T}$ at
$T\rightarrow \infty$. This  potential is also consistent with the
fact that there is no open string state at the end of the tachyon
condensation \cite{asen3}.

Now consider N non-BPS D-branes. They should be described
effectively by  non-abelian extension of the above action which
is \cite{mg1}\footnote{See \cite{bert}, for  the non-abelian
extension of Chern-Simons term for non-BPS D-branes.} \beqa
S&=&-\int
d^{p+1}\sigma \Tr\left(V(T)\sqrt{\det(Q^i{}_j)}\right.\labell{nonab}\\
&&\times\left.
e^{-\Phi(X)}\sqrt{-\det(P[E_{ab}(X)+E_{ai}(X)(Q^{-1}-\delta)^{ij}E_{jb}(X)]
+2\pi\alpha'F_{ab}+T_{ab})} \right)\,\,,\nonumber \eeqa where
$E_{\mu\nu}=g_{\mu\nu}+B_{\mu\nu}$. The  indices in this action
are raised  and lowered by $E^{ij}$ and $E_{ij}$, respectively.
The matrices $Q^i{}_j$ and $T_{ab}$ are  \beqa
Q^i{}_j&=&I\delta^i{}_j-\frac{i}{2\pi\alpha'}[X^i,X^k]E_{kj}(X)
-\frac{1}{2\pi\alpha'}[X^i,T][X^k,T]E_{kj}(X)\,\,,\labell{mq}\\
T_{ab}&=&2\pi\alpha'D_aTD_bT+D_aT[X^i,T](Q^{-1})_{ij}[X^j,T]
D_bT\nonumber\\
&&+iE_{ai}(X)(Q^{-1})^i{}_j[X^j,T]D_bT+iD_aT[X^i,T](Q^{-1})_i{}^jE_{jb}(X)
\nonumber\\
&&+iD_aX^i(Q^{-1})_{ij}[X^j,T]D_bT-iD_aT[X^i,T](Q^{-1})_{ij}
D_bX^j\,\,. \nonumber\eeqa The trace in the action \reef{nonab}
should be completely symmetric between all non-abelian expressions
of the form $F_{ab},D_aX^i,[X^i,X^j],D_aT,[X^i,T]$, individual
$T$ of the tachyon potential and individual $X^i$ of the Taylor
expansion of the closed string fields in the action\cite{mg5}.
The pull-back of closed string fields is defined in the static
gauge in which the derivatives are covariant derivative, \eg,
$P[\eta_{ab}]=\eta_{ab}+D_aX^iD_bX^j\eta_{ij}$. The gauge field
strength and covariant derivative of transverse scalars and
tachyons are\beqa
F_{ab}&=&\prt_aA_b-\prt_bA_a-i[A_a,A_b]\,,\nonumber\\
D_aX^i&=&\prt_aX^i-i[A_a,X^i]\,,\\
D_aT&=&\prt_aT-i[A_a,T]\,.\nonumber\eeqa Obviously the action
\reef{nonab} has $U(N)$ gauge symmetry when all branes are
coincident. This is resulted from the fact that all  gauge fields
corresponding to the open string stretched between branes are
massless in this case. When  all branes are separated in the
transverse space, only the gauge fields corresponding to the open
strings with both ends on one brane remain massless. In this case
the $U(N)$ symmetry breaks to $U(1)^N$ symmetry.

One may confirm various couplings in the action \reef{nonab} by
studying appropriate disk level S-matrix elements in string
theory \cite{mg1,mg2}. From  S-matrix elements point of view,
there is no difference between calculation of various fields
coupling on two non-BPS D-brane and on D-brane anti-D-brane. Both
involve exactly the same calculation. However, the difference is
only on the choice of Chan-Paton factors. For two non-BPS
D-branes, the Chan-Paton factors are such that the open string
fields are \beqa A_{a}=\pmatrix{A_a^{(11)}&A_a^{(12)}\cr
A_a^{(21)}&A_a^{(22)}},\,\,X^{i}=\pmatrix{X^{(11)i}&X^{(12)i}\cr
X^{(21)i}&X^{(22)i}},\,\,T=\pmatrix{T^{(11)}&T^{(12)}\cr
T^{(21)}&T^{(22)}}\,, \labell{M0}\eeqa where superscripts
$(11),(12),(21),(22)$ refers to the end of open strings, \eg
$(12)$ means the open string with one end on brane $1$ and the
other end on brane $2$. As we mentioned above, the symmetry of
the theory depends on the masses of $A_a^{(12)}$ and $A_a^{(21)}$
which can be calculated from the term $\Tr(D_aX^iD^aX_i)$ in the
action \reef{nonab}. They are massless for coincident branes
hence, the symmetry is $U(2)$, and they are massive for
non-coincident branes  hence, the symmetry is $U(1)\times U(1)$.

For D-brane anti-D-brane system, the matrices $A_a,X^i$ and $T$
are\beqa A_{a}=\pmatrix{A_a^{(1)}&0\cr
0&A_a^{(2)}},\,\,X^{i}=\pmatrix{X^{(1)i}&0\cr
0&X^{(2)i}},\,\,T=\pmatrix{0&\tau\cr \tau^*&0}\,. \labell{M1}\eeqa
We have changed the notation. The superscripts $(1)$ and $ (2)$
refer to the open string fields with both ends on brane $1$ and
$2$, respectively. $\tau (\tau^*)$ refers to the tachyon with one
end on brane $1(2)$ and the other end on brane $2(1)$. Since
there is no off-diagonal terms for the gauge field, the theory
has gauge symmetry $U(1)\times U(1)$ independent of the position
of  branes.  The above matrices satisfy  the following relations:
\beqa
&[X^i,X^j]=0,\,\,\,[X^i,A_a]=0,\,\,\,[A_a,A_b]=0,&\\
&[X^i,T]=\ell^i\pmatrix{0&\tau\cr -
\tau^*&0},\,\,\,[A_a,T]=(A_a^{(1)}-A_a^{(2)})\pmatrix{0&\tau\cr -
\tau^*&0}\,, &\nonumber \eeqa where $\ell^i=X^{(1)i}-X^{(2)i}$ is
the distance between the two branes. Hence, the non-abelian gauge
field strength and covariant derivative of the transverse scalars
become abelian, \ie $F_{ab}=\prt_aA_b-\prt_bA_a$ and
$D_aX^i=\prt_aX^i$. The matrix $Q^i{}_j$ in \reef{mq}  simplifies
to \beqa
Q^i{}_j&=&\left(I\delta^i{}_j+
\frac{|\tau|^2}{2\pi\alpha'}\ell^i\ell^kE_{kj}(X)\right)\,.\nonumber\eeqa
The inverse of this matrix is \beqa
(Q^{-1})^i{}_j&=&\left(I\delta^i{}_j-\frac{|\tau|^2}{(2\pi\alpha')\det(
Q) }\ell^i\ell^kE_{kj}(X)\right)\,, \eeqa where \beqa
\det(Q)&=&\left(I+\frac{|\tau|^2}{2\pi\alpha'}\ell^i\ell^kg_{ki}(X)\right)\,.\eeqa
It is easy to check that $Q^i{}_j(Q^{-1})^j{}_k=I\delta^i{}_k$.
The matrix $T_{ab}$ in \reef{mq} simplifies to \beqa
T_{ab}&=&(\det(Q))^{-1}\left(2\pi\alpha'D_aTD_bT
+i(E_{ai}(X)+\prt_a X^jE_{ji}(X))\ell^i\pmatrix{0&\tau\cr -
\tau^*&0}\
D_bT\right.\nonumber\\
&&\left.+iD_aT\pmatrix{0&\tau\cr - \tau^*&0}\
\ell^i(E_{ib}(X)-E_{ij}(X)\prt_bX^j)\right)\,.\labell{T1}\eeqa
Note that this matrix is not a real matrix, however, one expects
to have a real action after implementing the trace prescription.

Now using these facts that the trace in the action should be
symmetric,  $F_{ab},X^i,T^2$ are diagonal matrices and $D_aT,
[X^i,T]$ are off-diagonal matrices, one can implement the
symmetric trace by writing $D_aTD_bT$ and $[X^i,T]D_aT$ in
$T_{ab}$ in a symmetric form. That is,
$D_aTD_bT\rightarrow(D_aTD_bT+D_bTD_aT)/2$ and
$[X^i,T]D_aT\rightarrow ([X^i,T]D_aT+D_aT[X^i,T])/2$. Hence,
$T_{ab}$ becomes \beqa
T_{ab}&=&(\det(Q))^{-1}\left(\frac{I2\pi\alpha'}{2}
\left(D_a\tau (D_b\tau)^*+D_b\tau(D_a\tau)^*\right)\right.\\
&&\left.+\frac{i}{2}(E_{ai}(X)+\prt_a
X^jE_{ji}(X))\ell^i\left(\tau(D_b\tau)^*-\tau^*D_b\tau\right)\right.\nonumber\\
&&\left.+\frac{i}{2}\left(\tau (D_a\tau)^*-\tau^* D_a\tau\right)
\ell^i(E_{ib}(X)-E_{ij}(X)\prt_bX^j)\right)\,,\nonumber\eeqa where
the covariant derivative of tachyon is
$D_a\tau=\prt_a\tau-i(A_a^{(1)}-A_a^{(2)})\tau$. As anticipated
above, the matrix $T_{ab}$  is now real.

Now inserting the above expression in \reef{nonab} and performing
the trace, one finds the following action for $D_p\bar{D}_p$
system in arbitrary background at tree level: \beqa
S&\!\!\!=\!\!\!&-\int d^{p+1}\sigma
\left(\cV^{(1)}(|\tau|,\ell)e^{-\Phi(X^{(1)})}\sqrt{-\det
\textbf{A}^{(1)}}+\cV^{(2)}(|\tau|,\ell)e^{-\Phi(X^{(2)})}\sqrt{-\det
\textbf{A}^{(2)}}\right),\labell{action1}\eeqa where \beqa
\textbf{A}^{(n)}_{ab}&=&P^{(n)}[E_{ab}(X^{(n)})-\frac{|\tau|^2}
{2\pi\alpha'\det(Q^{(n)})}E_{ai}(X^{(n)})\ell^i\ell^jE_{jb}(X^{(n)})]+
2\pi\alpha'F_{ab}^{(n)}\nonumber\\
&&+\frac{1}{\det(Q^{(n)})}\left(\frac{2\pi\alpha'}{2}
\left(D_a\tau (D_b\tau)^*+D_b\tau(D_a\tau)^*\right)\right.\nonumber\\
&&\left.+\frac{i}{2}(E_{ai}(X^{(n)})+\prt_a
X^{(n)j}E_{ji}(X^{(n)}))\ell^i\left(\tau(D_b\tau)^*-\tau^*D_b\tau\right)\right.\labell{A1}\\
&&\left.+\frac{i}{2}\left(\tau (D_a\tau)^*-\tau^* D_a\tau\right)
\ell^i(E_{ib}(X^{(n)})-E_{ij}(X^{(n)})\prt_bX^{(n)j})\right)\,,\nonumber\eeqa
where $n=1,2$. In the above equation   $P^{(n)}[...]$ means
pull-back of closed string fields on the $n$-th brane, \eg
$P^{(1)}[\eta_{ab}]=\eta_{ab}+\prt_aX^{(1)}_i\prt_bX^{(1)}_j\eta^{ij}$.
The $D\bar{D}$ potential
 is \beqa
 \cV^{(n)}(|\tau|,\ell)&=&V(|\tau|)\sqrt{\det(Q^{(n)})}\nonumber\\
&=&V(|\tau|)\sqrt{1+\frac{|\tau|^2}{2\pi\alpha'}
\ell^i\ell^kg_{ki}(X^{(n)})}\,,\labell{tpot}\eeqa where
$V(|\tau|)$ is the tachyon potential of non-BPS D-brane. For
small $|\tau|$ it has the expansion \beqa
\cV^{(n)}(|\tau|,\ell)&=&T_p\left(1+\frac{2\pi\alpha'}{2}\left(\frac{\ell^i\ell^j
g_{ij}(X^{(n)}) }{(2\pi\alpha')^2}-
\frac{1}{2\alpha'}\right)|\tau|^2+{\cal O}(|\tau
|^{4})\right)\,.\eeqa The second term in the second parentheses
above  is the mass squared of the tachyon and the first term is
the mass squared of the string   stretched between  two branes,
\ie (tension)$^2\times$(length)$^2$. Note that potential had local
minimum at $|\tau|=0$ only when $\ell>\sqrt{2\pi^2\alpha'}$.

For trivial closed string background and for coincident branes,
$\ell=0$, the action \reef{action1} simplifies to \beqa S&=&-\int
d^{p+1}\sigma V(|\tau|) \left(\sqrt{-\det
\textbf{A}^{(1)}}+\sqrt{-\det
\textbf{A}^{(2)}}\right)\,,\labell{action2}\eeqa where \beqa
\textbf{A}_{ab}^{(n)}&=&\eta_{ab}+
\prt_aX^{(n)}_i\prt_bX^{(n)}_j\eta^{ij}+2\pi\alpha'F^{(n)}_{ab}+
\frac{2\pi\alpha'}{2}\left(D_a\tau(D_b\tau)^*+D_b\tau(D_a\tau)^*\right)\,.\eeqa
The action  \reef{action2} is the one proposed in \cite{sen3} when
two branes are coincident.

\section{$D_p\bar{D}_p$ interaction in open string channel}

In string theory, the interaction between two  D-branes with the
same RR charge is given by zero point function on cylindrical
world-sheet \cite{polch}. This world-sheet has two dual
descriptions. In terms of closed strings or in terms of open
strings. In the former which is a classical process, a closed
string is created by one brane. It propagates in the transverse
space between the two branes,  and then the other brane absorbs
it. In the latter which is a quantum process, a pair of open
strings stretching between the two branes are created by vacuum.
They propagate and then annihilate back to the  vacuum. The whole
amplitude  is zero due to an  identity.

This interaction can also be studied in low energy effective field
theories. If branes are far from each other, the only massless
fields are the graviton multiplete. So the  effective action is
in terms of gravity. That is to say, one of the branes produces
gravitational field due to its mass and RR charge, and
the other brane  moves in this background as a probe.
The repulsive force of RR charge cancels out the attractive force
of their mass. So the net classical force is zero.
On the other hand, when both branes are coincident, the lowest
mode of the open string stretching between the two branes and the
graviton multiplete are both massless. So the low energy
interaction can be either in terms of closed string or in terms of
open string fields. In the closed string channel, again the RR and
gravity forces cancel each other. In the open string channel, the
effective field theory includes  massless bosons and fermions.
The positive one loop vacuum energy of bosons cancels out the
negative vacuum energy of the fermions. Then, there is no quantum
interaction between the two branes which is consistent with the
gravity channel.

In string theory, when brane separation is larger than the string
length scale, interaction between a  $D_p$-brane and a
$\bar{D}_p$-brane  is again given by zero point function on
cylindrical world-sheet \cite{polch}. This world-sheet has again
two dual descriptions as in the $D_pD_p$ system. However, unlike
the $D_pD_p$ system,  there is no identity to make the whole
amplitude to be zero. This interaction can also be studied in a
low energy effective field theory. The only massless fields are
the graviton multiplete. So the low energy effective action is in
the closed string channel. Unlike the previous case, however, the
repulsive force of RR charge dose not cancel out the attractive
force of their mass. So the net classical force is non-zero, in
accord with string theory result.  Note that the separated
$D_p\bar{D}_p$ system is unstable, however, the decay mechanism is
not described by the above world-sheet nor by the above low energy
effective action.

On the other side, when brane separation is smaller than the
string length scale, interaction of  a  $D_p$-brane and a
$\bar{D}_p$-brane  in terms of cylindrical world-sheet gives a
complex force \cite{bank}. This indicates  there is a tachyonic
mode for the open string stretching between the two branes.
Moreover, it indicates that the force between the two branes in
full string theory is not given by the above simple world-sheet
any more. To follow the system in effective theory, one needs  an
effective potential which includes all order of tachyon fields.
Making use of the proposed effective action of \reef{action1}, we
would like to discuss in this section  the ${\it dynamics}$ and
the ${\it decay}$ of the system when brane separation is smaller
than the string length scale, and the ${\it decay}$ of the system
when brane separation is larger than the string scale.


To simplify the discussion, we consider  trivial background  and
consider  the case where branes are separated  in the $k$-th
transverse direction. Writing $\tau=\phi e^{i\theta}$ and
considering homogeneous case,   action \reef{action1} simplifies
to\beqa S&=&-\int d^{p+1}\sigma \cV(\phi,\ell) \left(\sqrt{-
\textbf{A}^{(1)}_{00}}+\sqrt{-
\textbf{A}^{(2)}_{00}}\right)\,,\labell{action3}\eeqa where \beqa
\cV(\phi,\ell)&=&V(\phi)\sqrt{1+\frac{\phi^2\ell^2}{2\pi\alpha'}}\,,\label{effpot}\\
\textbf{A}_{00}^{(n)}&=&-1+\dot{X}^{(n)i}\dot{X}^{(n)j}\eta_{ij}+
\frac{1}{1+\frac{\phi^2\ell^2}{2\pi\alpha'}}
\left(\dot{X}^{(n)k}\dot{X}^{(n)k}+2\pi\alpha'(\dot{\phi}^2+
\phi^2\dot{\theta}^2)\right)\,,\nonumber \eeqa where $i,j\neq k$.
Furthermore, writing $X^{(n)k}$ in terms of center of mass, $R$,
and in terms of brane separation, $\ell$, \ie $X^{(1)k}=R+\ell/2$
and $X^{(2)k}=R-\ell/2$, one finds that the action has no
potential for $R$,  $X^{(n)i}$ nor for $\theta$. So their
equations of motion  are satisfied for constant $R$, $X^{(n)i}$
and $\theta$. This simplifies the above action to \beqa S&=&-2\int
d^{p+1}\sigma\, \cV(\phi,\ell)
\sqrt{1-\frac{1}{1+\frac{\phi^2\ell^2}{2\pi\alpha'}}
\left(\frac{1}{4}\dot{\ell}^2+2\pi\alpha'\dot{\phi}^2\right)}\, .
\eeqa
Note that for $D_pD_p$ system, the corresponding action is
$S=-2T_p\int d^{p+1}\sigma \sqrt{1-\dot{\ell}^2/4}$ which gives
no potential for $\ell$. The effective force in this case is the
attractive Coulomb force due to the tension of the branes,
though, this force is canceled by the contribution from the
repulsive RR force.

The effective potential \reef{effpot} along  the $\ell$ direction
has a minimum at $\ell=0$ for any non-zero $\phi$. Because the
tachyon potential \reef{tacpo} goes to zero as
$e^{-\sqrt{\pi}\phi}$, the effective potential $\cV(\phi,\ell)$
along the $\phi$ direction has two minima   for
$\ell>\sqrt{2\pi^2\alpha'}$. One at $\phi=0$ and the other one at
$\phi\rightarrow\infty$. On the other hand,  for
$\ell<\sqrt{2\pi^2\alpha'}$, it has one maximum and one minimum
(see Fig1). Actually, these behaviours of the effective potential
$\cV(\phi,\ell)$ are independent of the form of tachyon potential
$V(\phi)$, so long as the tachyon potential goes to zero
asymptotically faster than $1/\phi$.

As an initial condition, consider a non-coincident stationary
$D_p\bar{D}_p$ system with a coherent quantum fluctuation for
tachyon, \ie $\phi=\epsilon$. The effective potential for this
initial condition is $\cV(\epsilon,\ell)\simeq T_p$. If
$\ell<\sqrt{2\pi^2\alpha'}$, the tachyon can roll down the
potential in  real time. This non-zero tachyon generates a time
dependent force for $\ell$. If $\phi_{cl}(t,\ell)$ is the
homogeneous tachyon rolling solution of the above action,  the
potential in the $\ell$ direction becomes, \beqa
\cV(\phi_{cl},\ell)&=&V(\phi_{cl}(t,\ell))\sqrt{1+
\frac{\phi_{cl}^2(t,\ell)\ell^2}{2\pi\alpha'}}\,,\eeqa which
produces a non-zero force. We speculate that this homogeneous
classical force might be the effective force between the two
branes.

On the other hand, if initial value of $\ell$ is
$\ell>\sqrt{2\pi^2\alpha'}$, there is no real time homogeneous
tachyon solution that penetrates the barrier of the tachyon
potential. This is consistent with the fact that the effective
force for ${\it dynamics}$ of  the branes in this case is in the
closed string channel, \ie the homogeneous Coulomb force.
However, the ${\it decay}$ of the branes should be in the open
string channel. In fact the  effective potential in Fig.1
indicates that there is an instanton effect where  system will
tunnel out of the false vacuum. There should be an inhomogeneous
 "bounce" solution to the effective action \cite{colman}.
This bounce solution should form  a bubble inside which the
tachyon is in the true vacuum and outside which the tachyon is in
the false vacuum. The solution should also include a throat
formation in the $\ell$ direction \cite{callen}. The classical
evolution  of the tachyon after penetrating through the potential
barrier,  should  make the bubble/throat to be expanded to
infinity leaving behind the true vacuum. Using the Euclidean
continuation of the following effective  action \beqa S&=&-2\int
d^{p+1}\sigma \cV(\phi,\ell)\sqrt{-\det\left(\eta_{ab}+\frac{1}{1+
\frac{\phi^2\ell^2}{2\pi\alpha'}}\left(\frac{1}{4}
\prt_a\ell\prt_b\ell+2\pi\alpha'\prt_a\phi\prt_b\phi\right)\right)},\nonumber\eeqa
one would find the "bounce" solution and  the  rate for decaying
$D_p\bar{D}_p$ to the closed string vacuum.  Similar studies have
been done in \cite{koji} using two-derivative truncation of the
BSFT effective action. We will leave this calculation for the
future works.

\begin{figure}
  \begin{center}
  \includegraphics[width=14cm]{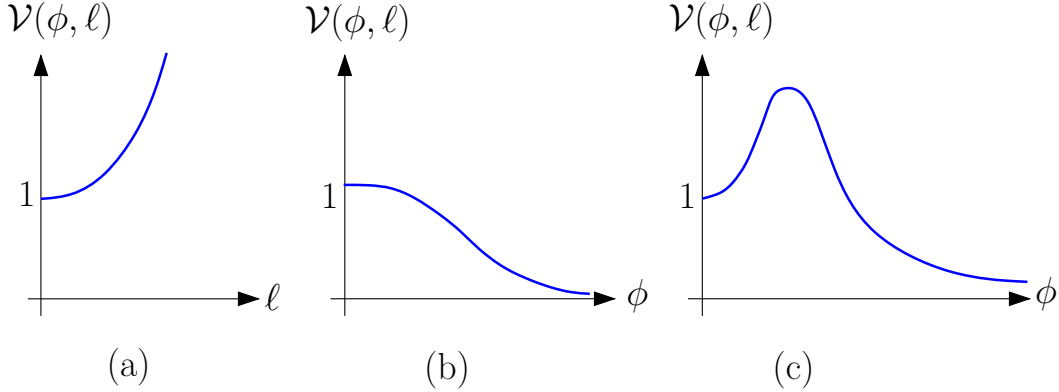}
  \end{center}

  \begin{quote}
  \caption{\it The D-branes anti-D-brane potential. a): For non-zero tachyon field, $\phi$,
  potential has  one minimum in the $\ell$ direction. b) For  brane separation $\ell<\sqrt{2\pi^2\alpha'}$,
  the potential has one maximum and one  minimum in the $\phi$ direction.
  In this case, there is a classical homogeneous tachyon rolling solution. c): For brane separation
  $\ell>\sqrt{2\pi^2\alpha'}$, the potential has two minimums in the $\phi$ direction. In this case, there
  is non-perturbative tachyon tunneling solution. The effective potential is symmetric under $\phi\rightarrow -\phi$ and
  $\ell\rightarrow -\ell$.}
  \end{quote}
  \label{fig:Fig1}
\end{figure}

Finally, one may couple the $D_3\bar{D}_3$ system to FRW gravity,
and studies its cosmological evolution. In the closed string
channel and for $\ell>>\sqrt{2\pi^2\alpha'}$, this cosmology  has
been studied in \cite{dvali}. In this study,  the motion of one
of the branes in the gravity background produced by the other
brane is considered as the cosmological evolution. In this case,
with $\ell$ as inflaton,  inflation occurs when branes are very
far from each other for which there is no decay of branes to
closed string vacuum. On the other hand in the open string
channel, if initially the value of $\ell=0$, one may consider
$\phi$ as inflaton. In this case, one  would find the cosmology of
the tachyon rolling \cite{Tachyonindustry}.
A problem for tachyon inflation is that inflation occurs  only
when tachyon is around the top of its potential. This does not
lead to enough number of e-folding. However, if initially the
value of $\ell\sim \sqrt{2\pi^2\alpha'}$,  the top of the tachyon
potential becomes more flat. This increases the number of
e-folding. For the case that the initial value of
$\ell>\sqrt{2\pi^2\alpha'}$, even the tachyon tunneling may cause
the inflation. In this case, the $D_3\bar{D}_3$ system
experiences the old inflation scenario \cite{old}. It would be
interesting to study this cosmologies in more details.


 {\bf Acknowledgement}: I would like to thank K. Hashimoto for
 discussion.


\end{document}